# Comparison among Classical, Probabilistic and Quantum Algorithms for Hamiltonian Cycle problem

**Giuseppe Corrente[1,2*], Carlo Vincenzo Stanzione[3,4] and Vittoria Stanzione[5]**

[1]Università di Torino – Computer Science Department, Corso Svizzera 185, Torino, Italy

[3]Università degli Studi di Pisa-Computer Engineering Department(DII), Via G. Caruso 16 - 56122 – Pisa, Italy

[4]Justonearth SRL-Via Beverino, 6 – 00168 Roma, Italy

[5]Università degli studi di Pisa-Physics Deparment E.Fermi, Largo Bruno Pontecorvo 3 - 56127 – Pisa, Italy

Corresponding author: Giuseppe Corrente[1]. Email : giuseppe.corrente@unito.it

**Abstract**: The Hamiltonian cycle problem (HCP), which is an NP-complete problem, consists of having a graph $G$ with $n$ nodes and $m$ edges and finding the path that connects each node exactly once. In this paper we compare some algorithms to solve a Hamiltonian cycle problem, using different models of computations and especially the probabilistic and quantum ones. Starting from the classical probabilistic approach of random walks, we take a step to the quantum direction by involving an ad hoc designed Quantum Turing Machine (QTM), which can be a useful conceptual project tool for quantum algorithms. Introducing several constraints to the graphs, our analysis leads to not-exponential speedup improvements to the best-known algorithms. In particular, the results are based on bounded degree graphs (graphs with nodes having a maximum number of edges) and graphs with the right limited number of nodes and edges to allow them to outperform the other algorithms.

## 1 Introduction

A Hamiltonian cycle is a cycle in an undirected or directed graph that visits each vertex exactly once. Hamiltonian Cycle problem (HCP) is the problem to determine if a Hamiltonian cycle exists in a given graph.

HCP is in NP, and precisely it is NP-complete. Shortly, the first proposition indicates that while it is almost impossible to find an efficient solution for HCP, it is straight in polynomial, often linear, complexity time to control if a given item is a solution or not. Obviously, generating and checking all possible solutions to find a correct one has exponential time complexity, and it is very difficult to obtain with little error margin a solution for them in another way, heuristics for example.

---

[1] Permanent Address: Via mercanti 6 , Torino, Italy

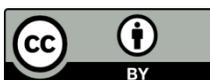

NP-complete problems are a class of decision problems within NP for which any other problem in NP can be reduced to them in polynomial time. In other words, if you could solve any NP-complete problem in polynomial time, you could solve all problems in NP in polynomial time. Therefore, a polynomial solution, more precisely an exact solution, given to HCP in its general formulation, would imply that $P = NP$, resolving one of the more discussed problems in Computer Science. Here we do not have a similar aim, but we want only to compare some algorithms to find the exact solution with the different computational models, which are the deterministic, probabilistic and quantum ones.

A simple deterministic solution to the HPC, for example, is the following: given a Graph with $N$ vertices, generate all permutation of its vertices which begin with 1, and this phase has complexity $O((n-1)!)$, where the ! is for the factorial function; check each permutation: meaning that we have to test if each vertex in the permutation is connected to the next by an edge, and that the last of the sequence is connected by an edge with that numbered 1; for each test we can assume complexity $O(n)$.

So, the whole complexity is $O(n(n-1)!) = O(n!)$ that can be approximated to $O\left(\sqrt{n}\left(\frac{n}{e}\right)^n\right)$ by the known Stirling formula [1].

In the simplest scenarios, that involve small graphs, this class of solutions is broadly used [2-3]. With larger graphs, though, a deterministic approach is computationally infeasible. Therefore, probabilistic and quantum algorithms come into play [4]; they do not guarantee finding a Hamiltonian cycle if one exists, but they can be effective in practice. In fact, the trade-off is between computational time, and the quality of the solution; these methods can provide reasonably good solutions within a reasonable time frame for many real-world instances of the HCP. The algorithms also differ by the class of graphs they are appliable to, because some of them are designed to give better solutions on HCP graph models with several constraints (such as the degree of nodes) than the ones that work on every graph. We will see some of these approaches in the Related Works section.

Overall, in this article we compare the most relevant algorithms and results between them and with others designed by ourselves, using HCP as a benchmark and a hint to compare different approaches and models of computation.

## 2 Related Works

We know [5] that for the sparse locally connected graphs, HCP is solvable in polynomial complexity time. Sparse locally connected graphs are graphs for which $|E| \leq 2|V| + k \, log_2|V|$, where $E$ are the edges, $V$ are the nodes or vertices and $k$ is the degree of each vertex, meaning the number of edges entering the node. Furthermore, if a graph $G$ is constructed so as to have $|V| = n$, with $n = m(k+1)$, where $m$ is the number of adjacent nodes and with the condition $m > 1$ and $k > 2$ integers, and if $G$ is $k - regular$, that is each node has degree $k$, then for $G$, HCP is in P as shown in [6].

In [7], they design a distributed algorithm that with high probability computes a Hamiltonian cycle in a random graph $G(n,p)$ with $p \geq (\log n)^{3/2}\sqrt{n}$, where $n$ is the number of the vertices and each possible edge occurs independently with probability $p$. They compute HCP with high probability for this graph class by using nodes as cooperative computing elements and reach the time complexity of $O(\log n)$. Obviously, they do not obtain this result for a generic graph $G$.

Another way to use probability in HCP solutions is to consider random walks in graphs. This type of research implies to construct a geometric polytope based on the graph to be examined, and using its extreme points, meaning points that do not lie in any open line segment joining two points, as starting point to analyze. Very positive experimental results are shown in [8] by using this type of approach.

In [9], they show a quantum computing model in which Hadamard gates are used to obtain all permutations of vertices in a superposition way. Then the Grover algorithm is applied to search for an eventual Hamiltonian cycle. So, in [9] the authors solve HCP for a general graph with high probability, but only a quadratic speedup is obtained compared to the deterministic model of searching classically among all possible solutions. The complexity time of this solution remains exponential.

As they show in [10], for an adiabatic quantum computing model, it is possible to use a particular methodology to obtain some code effectively testable on an existing prototype, the so-known D-Wave. They map the Hamiltonian cycle problem in QUBO problem, the mathematical algorithm for which D-Wave is designed. However, they don't observe an exponential speed-up over the number of nodes of the graph. They also underline problems with noise preparing and testing their examples.

Another quantum information technique for graph algorithms is quantum random walks. For example, in [11] quantum random walks are described for very particular graphs, such as hypercubes and a few tree-based graphs. However, using quantum walks for a generic graph was proved impossible [12].

Based on all these results, we reached the conclusion that very significant results in computational time are reached by restricting the graph classes on the problem instances.

The brute force algorithm presented in the introduction has a time-complexity of $O\left(\sqrt{n}\left(\frac{n}{e}\right)^n\right)$. For a generic graph, only some algorithms can outperform it. To introduce them, we also must consider that HCP can be reduced to the Traveling Salesman Problem (TSP). The latter is the problem of finding the minimum cost Hamiltonian cycle in a weighted graph. This is computed representing the graph as a cost matrix $W$, in which the $(i,j)$ element is the cost of the edge from node $i$ to node $j$. Then, it is possible to solve an HCP with an algorithm for TSP, putting in the $(i,j)$ element of $W$ the cost 1 if the edge from $i$ to $j$ exists, and 2 or more otherwise, and then verifying if the total cost of a TSP solution is equal to the number of nodes of the given graph, or not. Finally, we can apply Bellman-Held-Karp algorithm [13], which has complexity $O(2^n)$, or the algorithm designed in [14]. This latter is a quantum algorithm with complexity $O(1.728^n)$.

In addition, many heuristics and strategic approaches exist for TSP, such as Ant Colony Optimization Algorithm, Particle Swarm Optimization Algorithm, Artificial Bee Colony Algorithm and recently Genetic Algorithm [15-16], but they do not find the exact solution efficiently; they only search for one solution near to the optimal. Alternatively, also for TSP, as for HCP, quantum annealing algorithms have been proposed [17-19] but they still have too many limitations regarding noise and number of nodes of the graph that have to be elaborated [17].

In the following sections we will present our proposed probabilistic and quantum algorithms.

## 3 Methods and algorithms design

### 3.1 A probabilistic approach for HCP

Using a random walk approach on a graph $G$, we consider a memory-less path in which, if at the $t-th$ step we are at a node $v_t$, the probability to move to a neighbor is $\frac{1}{d(v_t)}$, where $d(v_t)$ is the degree of the vertex $v_t$. Obviously, the sequence of random nodes $(v_t : t = 1,2 ...)$ is a Markov chain. Now we add some conditions to the random walk for each single trial, as shown in the following sample of code. Let $n > 1$ be the number of the nodes of $G$.

Then we show:

1. *// Legend:*
2. *// Node is the class of every node*
3. *// $v_a \rightarrow$ property is a public member(property) of the node a*
4. *//$(v_a, v_b) == 1$ means that there is an edge between node a and b, $(v_a, v_b) == 0$ means there isn't*
5. *//$v_a \rightarrow$ visited == 1 means that the node a has already been visited*
6. *//$v_a \rightarrow$ numUnmarked is the number of unmarked(unvisited) neighbors of node a*
7. *//$v_a \rightarrow$ numNeighbors is the number of neighbors of node a*
8. *//$v_a \rightarrow$ neighbors is an array of pointers to every neighbor of node a*
9. *// chooseRandomNode() is the function that takes a node and returns the next one according to our approach*
10. *// getRandom() is a function that returns a random index in an array with a weight based on its values (in this // case the probability)*
11. 
12. Node chooseRandomNode(Node $v_t$){
13.     float probabilityArray[ $v_t \rightarrow$ numNeighbors];
14.     for (i=0; i< $v_t \rightarrow$ numNeighbors; i=i+1){
15.         if ($v_t \rightarrow$ neighbors[i] $\rightarrow$ visited ==0) {
16.             probabilityArray[i] = $\frac{1}{v_t \rightarrow numUnmarked}$;
17.         }
18.         else {
19.             probabilityArray[i] = 0;
20.         }
21.     }
22.     int randomIndex = getRandom(probabilityArray);
23.     return $v_t \rightarrow$ neighbors[randomIndex];
24. }
25. 
26. int main() {
27. 
28. for (t=1; (t<n && $v_t \rightarrow$ numUnmarked > 0) || (t==n && $(v_t, v_1)$==1); t=t+1){
29.     $v_t \rightarrow$ visited = 1;
30.     if (t==n && $(v_t,v_1)$==1){
31.         printf("$(v_1,v_2, ...., v_n, v_1)$ is an Hamiltonian cycle for G");
32.         break;
33.     }
34.     $v_{t+1}$= chooseRandomNode($v_t$);
35. }
36. 
37. return 0;

Follows:

**Theorem 1**. *Consider a random walk on a graph $G$ starting at node $v_1$ that follows the stop conditions stated in the above algorithm. Let $d(v_t)$ the degree of $v_t$, that is the number of vertices of $G$ adjacent to $v_t$. Let $m(v_t)$ be the number of vertices of $G$ adjacent to $v_t$, but already visited. Their difference is equal to the number of adjacent vertices to $v_t$ not visited yet (numUnmarked of $v_t$). Then, if the condition of Exit (line 30) is true then $(v_1, v_2, \ldots, v_n, v_1)$ is a Hamiltonian cycle for $G$ and its frequency, as the number of trials tends towards infinite, is:*

$$\frac{1}{\prod_{t=1}^{n-1}(d(v_t)-m(v_t))} . \tag{1}$$

*Proof.* The first part is simple to prove. If the condition in line 30 in the algorithm above is true, we have arrived at the $n-th$ step traversing all nodes different from each other. Furthermore, if there is an edge between $v_n$ and $v_1$, means that there is an edge between the vertex visited at step $n$ and the one visited at step 1 numbered 1 as given. So $(v_1, v_2, \ldots, v_n, v_1)$ is a Hamiltonian cycle.

We now prove the second statement of our theorem. It simply follows from the relative marking rule (line 29) of the algorithm above. The probability is calculated as product of independent events probabilities. Our algorithm's iteration steps depend on one another. Still, this dependence is already implicit in the term $m_t$ inside the probability designed in it, because at each step we are marking nodes, virtually increasing the term $m$ of some node. So, for this the probability that the second step is $v_1 v_2$ is $\frac{1}{d_1(d_2-m_2)}$ and the probability that at $t=3$ the visited nodes are $(v_1, v_2, v_3)$ is $\frac{1}{d_1(d_2-m_2)(d_3-m_3)}$. Note that $m_1$ is 0 because we haven't visited any nodes yet, so we write $d_1$ instead of $(d_1 - m_1)$. The probability that the random walk visit the sequence $(v_1, v_2, \ldots, v_n)$ is so already given by the final formula of the theorem, and if there is also an edge between the last visited node and the first it is certain that it is true also for the Hamiltonian cycle $(v_1, v_2, \ldots, v_n, v_1)$.

**Corollary 1**. *Considering the rate of success of our algorithm for each Hamiltonian cycle in the graph, and given that each random walk is $O(n)$ as time complexity, we can deduce that an upper bound of its expected time complexity is:*

$$O(n \prod_{i=1,2..n} d_i), \tag{2}$$

*where, for the sake of simplicity we chose $d(v_i) = d_i$.*

*Furthermore, if L is a lower bound to the number of Hamiltonian cycles of G in the case that G is Hamiltonian, then its expected time complexity is:*

$$O(\frac{n}{L} \prod_{i=1,2..n} d_i). \tag{3}$$

*Proof.* Given $m_i = 0$, considering that the complexity is the inverse of the success probability, and by considering the random walk from every node (hence the factor $n$), the first formula of the above corollary follows. The second formula instead follows from the first and the corrected hit rate, based on a supposed lower bound of the number of Hamiltonian cycles in the graph, which allows us to increase the probability of finding one, therefore improving the time complexity.

We have designed in such way a probabilistic algorithm that, with high probability, solves for a generic graph $G$ with a better time complexity than classical brute-force deterministic algorithm presented in the introduction assuming simply that:

$$d_i < \frac{n}{e} \quad \text{for } i = 1,2\ldots n. \tag{4}$$

However, we remain in an exponential time complexity.

Comparing our algorithm to the quantum computational search algorithm shown in [9], we also can do better but only with the following stronger assumption:

$$d_i < \sqrt{\frac{n}{e}} \quad \text{for } i = 1,2\ldots n. \tag{5}$$

We are not discussing here the advantage of decreasing the degree from $d_i$ to $d_i - m_i$ during the random walk. Consequently, our study of time complexity is valid also for perfectly Markov random walks, not based on previous choices.

Furthermore, to outperform respectively the best-known classical [13] and quantum [14] algorithm for HCP, the following conditions must be true:

$$\prod_{i=1,2..n} d_i \;<\; 2^n; \tag{6}$$

$$\prod_{i=1,2..n} d_i \;<\; (1.728)^n. \tag{7}$$

### *3.2 The quantum version of our probabilistic algorithm*

In this chapter, we show a few quantum versions of our algorithms in a QTM format, as we already did for some known algorithm in [20].

We give the following definition of a Quantum Turing Machine [20-21].

A Quantum Turing Machine (QTM for short) $M$ can be seen as the quantum version of a Turing Machine, usually described by the 7-tuple $M = (Q, \Sigma, \Gamma, \delta, q0, \Box, F)$ with a condition of a unitary evolution, where:

- $Q$ is the (finite) set of internal states $\{q_i \mid i \in N,$ and $q_i$ is usually referred to as the current state$\}$;
- $\Sigma$ is the input alphabet, $\Gamma$ is the finite set of symbols called tape alphabet - a sequence of cells containing symbols (one in each cell)- (i.e., $\Gamma \cup \Box$) and it usually contains at least 1-used to code natural numbers in unitary notation- and $\Box$, the blank symbol;
- $\delta : \Gamma \times Q \times \Gamma \times Q \times \{L, N, R\} \to \mathbb{C}_{[0,1]}$ is the transition function that allows to move from one state to another, giving the amplitude of each step. The square of this function represents the probability of having that step if a measurement occurs. Furthermore, we have the condition $\delta \in \mathbb{C}$ and $|\delta| \leq 1$. $L, N, R$ are allowed moves of the tape head, respectively Left, None - no allowed head move - and Right;
- $q_0$ ( a distinguished member of $Q$) is the initial state;
- $F$ ( a subset of $Q$) is the set of final states (one final state is sufficient).

Here we itemize other relevant features:

- A tape is a pair of strings $w_L$ and $w_R$ such that $w_L \in \Box^\infty \Gamma^*$ and $w_R \in \Gamma^* \Box^\infty$;
- $h\,(head) \in \Gamma$ is the head of the tape whenever is the rightmost symbol of $w_L$;
- A configuration of $M$ is a triple in $Q \times (\Box^\infty \Gamma^*) \times (\Gamma^* \Box^\infty)$;
- The initial configuration is represented as $<q_0, \Box^\infty w, \Box^\infty>$ where $w \in \Gamma^*$ is the input;
- A final configuration is $<q_F, \Box^\infty w, \Box^\infty>$ where $q_F \in F$ and $w \in \Gamma^*$ is the output; we assume that if a final configuration is a superposition of more than one, then all of them are in a final state.

A QTM-computation [15] is a (finite) set of configurations $C_M$ above which $\delta$ determines a mapping a: $C_M \times C_M \rightarrow C_{[0,1]}$ such that for each $c_1, c_2 \in C_M \times C_M$, $a(c_1, c_2) \in C_{[0,1]}$ represents the amplitude of the transition of $M$ from $c_1$ to $c_2$. This matrix must be unitary.

Consequently, for each configuration $c_0$ and all its next configurations $c_1, \ldots, c_k$, if $\alpha_i$ is the amplitude representing the transition from $c_0$ to a generic configuration $c_i$, then:

$$\sum_{i=1}^{k}|\alpha_i|^2 = 1, \tag{8}$$

where $|\alpha_i|^2$ represents the probability of going from $c_0$ to $c_1$, but all the configurations $c_1, \ldots, c_k$ occur in a parallel way - step by step- until a measurement is effectuated.

Note that after the first step, the starting configuration $c_0$ can be also a superposition of configurations. In this case, the next configuration is determined by the transition function as well but weighting each component of $c_0$ with the relative amplitude.

We now want to give a quantum version of our algorithm.
We start with the definition of the following QTM:

- $Q = \{\Phi_0, \Phi_1, \Phi_2, \Phi_3 \ldots, \Phi_n, \Phi_F\};$ \hfill (9)

- $\Sigma = \{\{1,2,3 \ldots, n, H, NH\} \cup \{\square\}\}^*$, where $* = \{1,2,\ldots, n+1\}$ and $n > 1;$ \hfill (10)

- $q_0 = \Phi_0;$ \hfill (11)

- $F = \{\Phi_F\};$ \hfill (12)

Each symbol refers to the previous generic definition of a QTM, $H$ and $NH$ represent respectively the symbols for a found Hamiltonian cycle and a not found one, $\square$ is the blank symbol and $n$ is the number of the nodes of a given graph $G$.

Note that with this definition each cell of the tape may contain not only a single symbol and their superposition, but also partial or full paths of the graph and their superposition.

Then, we define $\delta$ with the following rules:

$$\delta\ (\square, \Phi_0, 1, \Phi_1, R) = 1, \tag{13}$$

$$\delta\ (\square, \Phi_1, f(1), \Phi_2, R) = 1, \tag{14}$$

$$\delta\ (\square, \Phi_2, f(1f(1)), \Phi_3, R) = 1, \tag{15}$$

$$\delta\ (\square, \Phi_3, f(1f(1)f(1, f(1)), \Phi_4, R) = 1, \tag{16}$$

$$\delta\left(w_L NH \boxdot, \Phi_3, NH, \Phi_4, R\right) = 1, \tag{17}$$

where $f$ is a function of the given graph $G$ and is supposed to be implemented in a way that we can go easily from a node to its neighbors, we have a QTM that represents the quantum version of our algorithm.

Let $w_L$ be the string contained at step $k - th$ in the input tape at the left of the cursor, then we have:

$$\delta\left(w_L \boxdot, \Phi_k, f(w_L), \Phi_{k+1}, R\right) = 1 \; for \; k = 1, 2, \ldots n-1, \tag{18}$$

$$\delta\left(w_L NH, \Phi_k, NH, \Phi_{k+1}, R\right) = 1 \; for \; k = 1, 2, \ldots n-1. \tag{19}$$

And finally:

$$\delta\left(w_L \boxdot, \Phi_n, f(w_L), \Phi_F, R\right) = 1. \tag{20}$$

$$\delta\left(w_L NH, \Phi_n, NH, \Phi_F, R\right) = 1. \tag{21}$$

In (20) $f$ returns $H$ only if $w_L$ is a Hamiltonian cycle, otherwise $f$ returns $NH$.

Now a measure on the input tape, precisely in the cell relative to the cursor position, that is the $n + 1 - th$ cell from origin position, is done.

Function $f$ is an oracle, a function that has access to the graph $G$ representation in an efficient way. Initially, it starts from node 1 of $G$. It returns a superposition of paths of length 2 ending with 1 neighbor. The cursor moves always right. The oracle function considers all information written in the tape, also because it is summarized at the rightmost position of the written part of the tape- in the last cell exactly. If possible, $f$ without revisiting nodes, increments each path of the superposition of next cell node.

When it is not possible, $f$ writes the $NH$ symbol on the tape. This symbol is eventually propagated for all remaining steps of the path. At the last step, which is always the $n + 1 - th$ step, $f$ determines if there is a Hamilton cycle, that is a $n$ length path and an edge to node 1 closing the cycle, and writes for that path the symbol $H$. Otherwise $f$ writes $NH$.

The probability of finding $H$ as a result, given the previously defined $f$, is greater than zero only if the graph is Hamiltonian. In this case, a read operation on the whole tape would result in a Hamiltonian cycle.

**Theorem 2**. The probability of a hit result in the final measure for the above quantum algorithm is:

$$\frac{1}{\prod_{t=1}^{n-1}(d(v_t) - m(v_t))}. \tag{22}$$

*Proof*. Indeed, the time of evolution of our algorithm, if we exclude the latest step, the measurement, is exactly $n + 1$ steps. But we must consider that all results are processed in the number of repeated trials to have a good probability to measure an $H$ result if the graph is Hamiltonian. Also, we must consider that if we measured only the single last cell of the tape, it can have only two symbols, $H$ or $NH$, because the $H$ symbol may be the final symbol of a single or few paths, while $NH$ the final symbol of a lot of paths processed. The number of paths processed is the same as in our probabilistic algorithm of the previous section. This quantum algorithm reproduces exactly the probabilistic one of the previous section. Indeed, we used the quantum framework to obtain the same functionality and we didn't gain an advantage. With this we demonstrate the Theorem 2. In addition, note that in each of the $n$ steps of the processing algorithm we call the function $f$, which we supposed to be linear and however not less than constant as time complexity. In order to make a new measure each trial must run fully and independently. Two corollaries follow.

**Corollary 2**. The complexity of the above quantum algorithm derives from the repeated number of its executions and measure operations. This must be proportional to the inverse of the probability given in Theorem 2.

**Corollary 3**. The complexity of the probabilistic and quantum algorithms of Theorem 1 and Theorem 2 respectively are the same.

This follows from what is stated in [22], when quantum computation is not used with all or a large part of its features, often the respective algorithms don't perform computationally faster than probabilistic ones for a similar problem. More in general, we can say that in order to obtain a quantum advantage, it is necessary to rely on purely quantum features rather than trying to replicate a classical version of our problem within a quantum framework.

### 3.3 A quantum interference-based algorithm for HCP

If we want a better result, we may relax some QTM conditions and admit the following modification to the last rules (20) and (21), substituting them with the following ones:

$$\delta(w_L \square, \Phi_n, f(w_L), \Phi_F, R) = -1^{RND(0,1)}, \qquad \text{if head}(f(w_L)) = NH, \tag{23}$$

$$\delta(w_L \square, \Phi_n, f(w_L), \Phi_F, R) = 1 \qquad \text{if head}(f(w_L)) = H, \tag{24}$$

$$\delta(w_L NH, \Phi_n, NH, \Phi_F, R) = -1^{RND(0,1)}, \tag{25}$$

where $RND(0,1)$ is a random extraction of 0 or 1 both with 50% of probability, and with head a function that reads the rightmost position.

Due to (23), (24), (25), if we read the symbol $NH$ in the state $\Phi_n$ in a right movement, the transition to another state with the symbol $NH$ has amplitude $+1$ or $-1$ both with probability 0.5. So, it is not a standard QTM anymore, but it is a theoretically valid extension of the original QTM, also because it induces again a unitary transformation on QTM configurations. We obtained a probabilistic QTM that respects the unitary constraint and whose associated amplitudes are functions of the status of the tape and of the state of the machine.

The presence of negative interference allows us to access more paths and, if the graph is Hamiltonian, the result of measurement is $H$ with higher probability. Consequently, the complexity of our algorithm is better than the one from the previous algorithm version.

**Theorem 3**. The time complexity of the algorithm with the new rules (23), (24) and (25) instead of (20) and (21) is:

$$O\left(n\sqrt{\prod_{i=1}^{n-1} d_i}\right). \tag{26}$$

*Proof.* The *NH* results are affected by negative interference with each other, but in a random way. So, the final weight may be seen as the sum of negative and positive weights in a fully random modality. This behaviour can be described as the equivalent of a particle in a Brownian motion: we have as minimum and maximum expected values after $t$ steps, each step consisting of an increment or a decrement of one unit of:

$$\mp \sqrt{\frac{2t}{\pi}}. \tag{27}$$

Consequently, we can derive (26) from our definition of $d_t$ in section 3.1.

So, we have a quadratic speedup, the same obtained applying to our algorithm a quantum search of Grover type [22-24], for example in a similar manner that in [9].

Now we explain better the oracle function of our quantum algorithm.

Note that, given that $(i, j) \in E$, where $j$ is a neighbor of $i$ which has not been visited yet, and $i$ is the last node visited of the current partial path, we have:

$$f(w_L) = w_L \sum_j \sqrt{\left(\frac{1}{(d_i - m_i)}\right)} |j\rangle, \tag{28}$$

where, as in the non-quantum case, $d_i$ is the degree of the node $i$, and $m_i$ is the number of the already visited neighbors of $i$ in the current path. If $w_L$ contains a partial path, and the last node of it has more than one not visited nodes, then $w_L$ is incremented in a superposition way of one node in the next cell.

If all the neighbors of $i$ have been visited, and if the length of the path is less then $n$ (number of G nodes) or there is no edge $(i, l)$, the next symbol added to the path is *NH*. Otherwise, if the length of the current path is $n$, and such an edge exists, then the symbol *H* is added to the right of the path. At the end, all superposition paths have length $n + 1$ and terminate with *H* or *NH* and the necessary steps to evaluate the graph are exactly $n + 1$ in all cases. The final states in each superposition are reached at the $n + 1 - th$ step. If the graph is Hamiltonian, the last symbol of the tape is a superposition between *H* and *NH*. Otherwise, if the examined graph is not Hamiltonian the last symbol will be *NH* with no superposition.

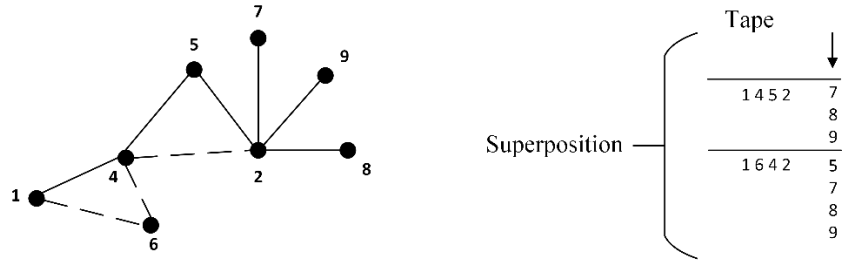

**Figure 1:** An example of superposition of two visited paths in a part of the graph G and on the tape, and of the next node choice. Three possible choices are added to the first path and four to the second. The resulting superposition of seven paths is the result of this choice.

On the left-hand side of Figure 1, we can see a graph $G$ with two partial paths in superposition. We suppose (1,4), (1,6), (4,6), (2,4), (2,5), (2,7), (2,8), (2,9), (4,5) belong to $E$, which are the edges of the graph G. If we suppose to be at the fourth walking edge of our algorithm, some superposition path has been generated. Two paths in superstition are 1,4,5,2 and 1,4,2,5, but for this second one, the next symbol associated to it will be deterministically NH, because from 5 we can go only to node 4, an already visited node. Two other paths in superposition are 1,6,4,2 and 1,6,4,5 and both have the possibility to go to a not visited neighbor, respectively 5 and 2 for example.

On the right-hand side of Figure 1, we consider the two possible partial paths: 1,4,5,2 and the 1,6,4,2. For the first one, the next choice can be 7,8,or 9, each with amplitude $\sqrt{\frac{1}{3}}$; while for the second path, the choice for the next node can be 5, 7, 8 or 9, each with amplitude $\sqrt{\frac{1}{4}}$.

All these paths are possible, then they are in superposition. The chosen one will be overwritten on the tape, each weighted with their appropriate amplitude. At each phase, the path length of all superposition paths will be the same. If a path terminates before visiting all possible nodes, from the rules derives that the terminating positions will be filled with the $NH$ symbol.

The oracle function $f$ determines next node together with the amplitude of each choice too.

## 4 Results

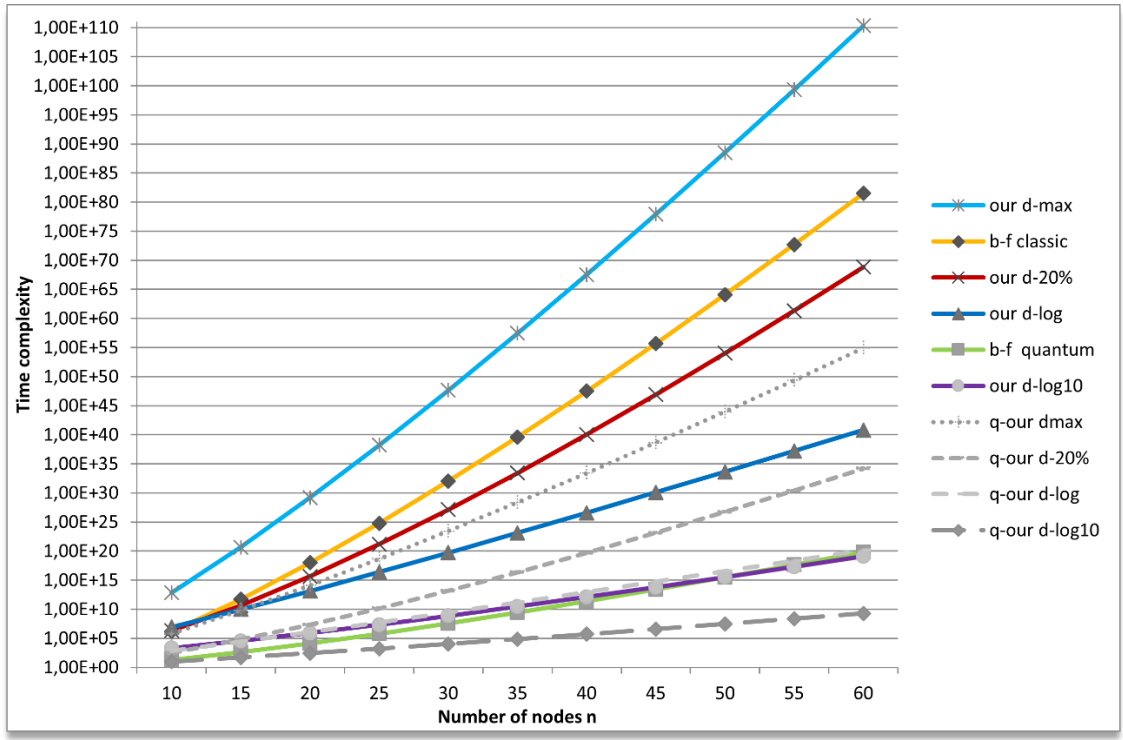

**Figure 2:** Comparison between our probabilistic algorithm and different quantum and brute force algorithms

**Table 1:** Final short summary of different algorithms (or the same ones applied to different classes of graph) compared in order of time complexity, for n that tends to infinity

| Algorithm | Maximum degree of G nodes | Strategy/type | Time complexity | Observations |
|---|---|---|---|---|
| our d-max | n-1 | Brute force/probabilistic | $O(n \prod_{i=1,2..n} d_i)$ | Valid only on graphs with nodes with a maximum degree of n-1, computational time infeasible for big graphs |
| b-f classic | Independent | Brute force/deterministic | $O\left(\sqrt{n}\left(\frac{n}{e}\right)^n\right)$ | Exact solutions*, but computational time infeasible for big graphs |
| our d-20% | n/5 | Brute force/probabilistic | $O(n \prod_{i=1,2..n} d_i)$ | Valid only on graphs with nodes with a maximum degree of n/5-degree, better than all brute force deterministic algorithms |
| our d-log | Ln(n) (log base e) | Brute force/probabilistic | $O(n \prod_{i=1,2..n} d_i)$ | Valid only on graphs with nodes with a maximum degree of ln(n), better than all brute force deterministic algorithms |
| b-f quantum | Independent | Brute force/quantum (with Grover search) | $O\left(n\left(\frac{n}{e}\right)^{\frac{n-1}{2}}\right)$ | Valid for every graph, better than all brute force deterministic algorithms |
| our d-log10 | Log(n) (log base 10) | Brute force/probabilistic | $O(n \prod_{i=1,2..n} d_i)$ | Valid only on graphs with nodes with a maximum degree of log(n), better than all brute force deterministic algorithms and better than *b-f quantum* starting from n=60 |
| best cl | Independent | Deterministic (best known) | $O(2^n)$ | Exact solutions*, best deterministic algorithm known |
| best q | Independent | Quantum (best known) | $O(1.728^n)$ | Valid for every graph, best quantum algorithm known |
| q-our d-max  q-our d-20%  q-our d-log  q-our d-log10 | n-1  n/5  Ln(n) (log base e)  Log(n) (log base 10) | Quantum interference/ probabilistic (extended QTM) | $O(n \sqrt{\prod_{i=1,2..n} d_i})$ | If condition 31 is verified, the algorithm is better than the others for every graph |

*Obviously deterministic algorithms are the only ones that can solve the HCP with 100% probability.

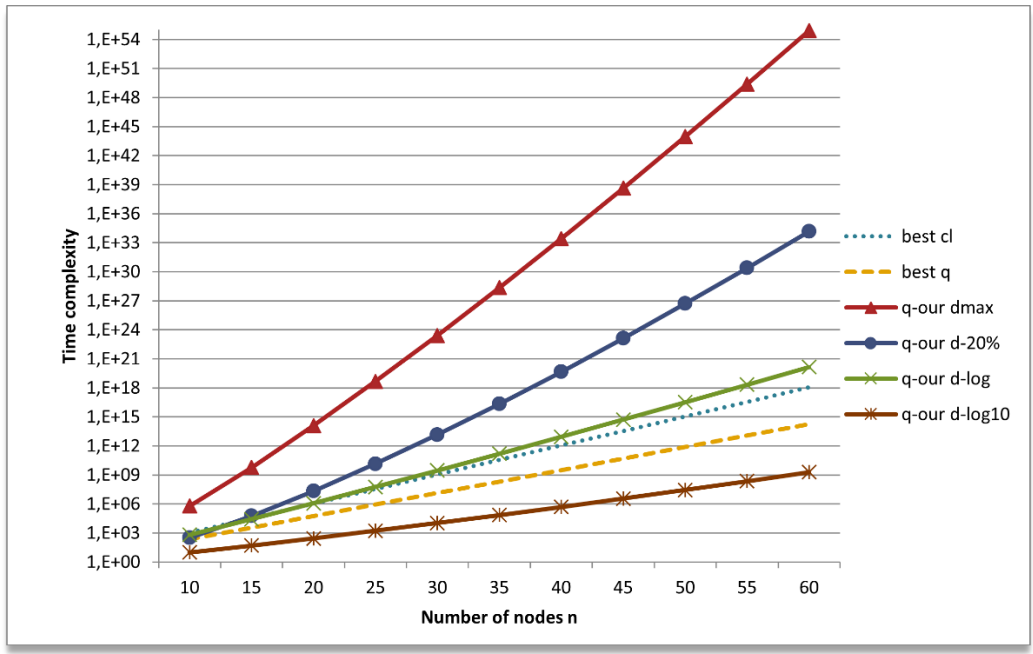

**Figure 3:** Comparison between our probabilistic quantum algorithm and the best-known classic and quantum ones

In Figure 2, we show an overview of all the algorithms, classic, probabilistic and quantum ones, shown schematically in Table 1. On the y-axis we represent the expected execution time steps in logarithmic scale. On the x-axis we represent the number of nodes. The first thing that catches the eye is that the classical algorithms can all be over-performed in terms of time complexity by the quantum ones. The brute force classical algorithm (*b-f classic*) is better than our classical algorithm only for the maximum degree graph class; while obviously the brute force quantum algorithm (*b-f quantum*) is always better than *b-f classic* labelled algorithm, this latter being the brute force algorithm that analyzes classically all the possible permutations of the G vertices. *B-f quantum* is a quantum search-based algorithm that works on all possible permutations; it can be over-performed by our probabilistic algorithm applied to graphs whose degree is limited by the base-ten logarithm of the number of nodes (*our d-log10*) and obviously by our quantum version with the same graphs class (*q-our d-log10*).

Following the same criteria, with the labels *our d-20%* and *our d-log*, we represent time complexity of our algorithm with parameter d, the maximum degree of the graph nodes, given as limited respectively to the 20% and to the logarithm of nodes number. The respective quantum versions of these algorithms are given by *q-our d-20%* and *q-our d-log*. Note that all these algorithms have an exponential time complexity and that only comparing them while they operate on the same classes of graph is correct.

In Figure 3, we compare our algorithms with best classical and best quantum known algorithms, respectively indicated with labels *best cl* and *best q*. These are over-performed only by the quantum version of our algorithm. Indeed, Figure 2 and Figure 3 show that only our quantum algorithm applied to base-ten logarithm limited degree graph class, labelled *q-our d-log10*, among those examined, is better than best classical and quantum known algorithms.

Moreover, if we limit $n$ to a few hundreds, our quantum algorithm can achieve the same time complexity of *q-our d-log10* for every class of graph following this relation:

$$\prod_{i=1,2..n} d_i \;\; < \;\; (Log\, n)^n \;, \tag{29}$$

while our algorithm performs better than best known algorithms also for $n$ that tends to infinity if

$$\prod_{i=1,2..n} d_i \;\; < \;\; (1.728)^{2n}, \tag{30}$$

which can be written as:

$$\prod_{i=1,2..n} d_i \;\; < \;\; (2.986)^n \;. \tag{31}$$

## 5 Conclusions

We have designed and analyzed a probabilistic algorithm for a Hamiltonian Cycle Problem (HCP), even if without obtaining for the general case a relevant improvement. We also designed two quantum versions of our algorithm, respectively without and with negative interference added. We compared them, the latter in particular, with a classical and a quantum algorithm. In this case, for some classes of graphs, for example, the upper bounded degree graphs, with the assumption of the validity of (31), our algorithm performs better. It outperforms not only the quantum algorithm based on Grover searches on all permutations, but also the best classical and quantum already known algorithms for HCP.

Anyway, we can obtain a quadratic speedup with our quantum algorithm with respect to the classic version of it, only if we use also quantum negative interference and not only quantum superposition.

An interesting aspect arising from this work is the possibility of comparing probabilistic with quantum computational models. Indeed, this work can also be seen as a hybrid approach combining quantum and probabilistic elements. Also, it is worth deepening our knowledge of our version of a QTM, as it is a valid tool for algorithm design. Further studies should be done to compare our algorithm with that written specifically for graphs of bounded average degree, for example those in [25-26].

Another theme that emerges from this work is the exploration of other NP problems and determining under what conditions, with constraints on the original problem or extensions of the quantum model, they become tractable.

**Acknowledgement:** none.

**Funding Statement:** The authors received no specific funding for this study.

**Author Contributions:** The authors confirm contribution to the paper as follows: study conception and design: Giuseppe Corrente; algorithms design and refinement: Giuseppe Corrente, Carlo Vincenzo Stanzione, Vittoria Stanzione; analysis and interpretation of results: Giuseppe Corrente, Carlo Vincenzo Stanzione, Vittoria Stanzione; draft manuscript preparation: Giuseppe Corrente, Carlo Vincenzo Stanzione, Vittoria Stanzione. All authors reviewed the results and approved the final version of the manuscript.

**Availability of Data and Materials:** Not applicable.

**Conflicts of Interest:** The authors declare that they have no conflicts of interest to report regarding the present study.

**Bibliography**

[1]     Dutka, Jacques "The early history of the factorial function", Archive for History of Exact Sciences, vol. 43, pp. 225–249, 1991.

[2]     Eric Lewin Altschuler Martin Lades Richard Stong, "Finding Hamiltonian Cycles", Science, vol. 273, pp. 413-415, 1996.

[3]     P. Baniasadi, V. Ejov, J.A. Filar, M. Haythorpe and S. Rossomakhine. "Deterministic "Snakes and Ladders" Heuristic for the Hamiltonian cycle problem", Mathematical Programming Computation, vol. 6, pp. 55–75, 2014.

[4]     Anuradha Mahasinghe, Richard Hua, Michael J. Dinneen, and Rajni Goyal "Solving the Hamiltonian Cycle Problem using a Quantum Computer" In Proceedings of the Australasian Computer Science Week Multiconference (ACSW '19). Association for Computing Machinery, New York, NY, USA, Article 8, 1–9, 2019

[5]     A. S. van Aardt, A. P. Burger, M. Frick, C. Thomassen and P. de Wet, "Hamilton cycles in sparse locally connected graphs," Discrete Applied Mathematics, vol. 257, pp. 276-288, 2019.

[6]     M. Haythorpe, "On the Minimum Number of Hamiltonian Cycles in Regular Graphs," *EXPERIMENTAL MATHEMATICS,* vol. 27, no. 4, pp. 426-430, 2018.

[7]     V. Turau, "A distributed algorithm for finding Hamiltonian cycles in Random Graphs in O(log n) Time," *Theoretical Computer Science,* vol. 846, pp. 61-74, 2020.

[8]     T. Kalinowski and S. Mohammadian, "Feasible bases for a polytope related to the Hamilton cycle problem," *arXiv preprint arXiv:1907.12691,* 2019.

[9]     V. C. Raj and M. S. Shivakumar, "Applying Quantum Algorithm to Speed Up the Solution of Hamiltonian Cycle Problems," in *IFIP TC12 International Conference on Intelligent Information Processing*, Springer: Boston, vol. 228, pp. 53-61, 2006.

[10]    A. Mahasinghe, R. Hua, M. J. Dinneen and R. Goyal, "Solving the Hamiltonian Cycle Problem using a Quantum Computer," in *Proceedings of the Australasian Computer Science Week Multiconference*, pp. 1-9, 2019. https://doi.org/10.1145/3290688.3290703

[11]    S. E. Venegas-Andraca, "Quantum walks: a comprehensive review," *Quantum Information Processing,* vol. 11, p. 1015–1106, 2012.

[12]    A. Glos and J. A. Miszczak, "The role of quantum correlations in Cop and Robber game," *Quantum Studies: Mathematics and Foundations,* vol. 6, no. 1, pp. 15-26, 2019.

[13]    M. Held and R. Karp, "A dynamic programming approach to sequencing problems," *Journal of SIAM,* vol. 10, pp. 196-210, 1962.

[14]    A. Amabainis, K. Balodis, J. Iraids, M. Kokainis, K. Prusis and J. Vihrovs, "Quantum Speedups for Exponential-Time Dynamic Programming Algorithms," 2018. https://doi.org/10.48550/arXiv.1807.05209

[15]    T. Kaur, A. K. Lamba and R. Talwar, "A Review: Optimized Solutions for Travelling Sales Person Problem," *International Journal of Engineering Trends and Applications (IJETA),* vol. 8, no. 3, pp. 36-40, 2021.

[16]    C. Dahiya and S. Sangwan, "Literature Review on Travelling Salesman Problem," *International Journal of Research,* vol. 5, no. 16, pp. 1152-1155, 2018.

[17]    S. Jain, "Solving the Traveling Salesman Problem on the D-Wave quantum computer," *Frontiers in Physics - Quantum Engineering and Technology,* 2021. https://doi.org/10.3389/fphy.2021.760783

[18]    Warren, R.H. "Solving the traveling salesman problem on a quantum annealer", SN Appl. Sci. 2, 75 (2020).

[19]    Stogiannos, E.; Papalitsas, C.; Andronikos, T. "Experimental Analysis of Quantum Annealers and Hybrid Solvers Using Benchmark Optimization Problems" Mathematics, 10, 1294, 2022.

[20]    G. Corrente, "Translation of Quantum Circuits into Quantum Turing Machines for Deutsch and Deutsch-Jozsa Problems}," *Journal of Quantum Computing,* vol. 2, no. 3, pp. 137-145, 2020.

[21]    Guerrini, S.; Martini, S.; Masini, A. "Quantum Turing Machines: Computations and Measurements", *Appl. Sci.*, 10, 5551,2020.

[22]     J. Gruska, Quantum Computing, McGraw-Hill, 1999.

[23]     G. Corrente, "Reflections on probabilistic compared to quantum computational devices," *International Journal of Parallel, Emergent and Distributed Systems,* pp. 251-261, 2020. https://doi.org/10.1080/17445760.2020.1805610

[24]     L. Grover, "A fast quantum mechanical algorithm for database search," in *Proceedings of the twenty-eighth annual ACM symposium on Theory of computing*, pp.212-219, 1996. https://doi.org/10.1145/237814.237866

[25]     M. Cygan and M. Pilipczuk, "Faster exponential-time algorithms in graphs of bounded degree," *Information and Computation,* vol. 243, pp. 75-85, 2015.

[26]     Moran Feldman, " Algorithms for Big Data", The Open University of Israel, https://doi.org/10.1142/11398, pp. 267-307, 2020.